\documentclass[journal, 10pt]{IEEEtran}
\IEEEoverridecommandlockouts\IEEEpubid{\makebox[\columnwidth]{}\hspace{\columnsep}\makebox[\columnwidth]{ }}
\usepackage[utf8]{inputenc}
\usepackage[english]{babel}
\usepackage{amsmath, bm}
\usepackage{graphicx}
\usepackage{dsfont}
\usepackage{indentfirst}
\usepackage{acronym}
\usepackage{xcolor}
\usepackage{cite}
\usepackage{tikz}
\usepackage{listings}
\usepackage{subfigure}
\usepackage{amssymb}
\usepackage{cancel}
\usepackage{tabularx}
\usepackage{hyperref}
\usepackage{amsthm}
\usepackage{url}

\newtheorem{theorem}{Theorem}[section]

\begin{document}
\definecolor{codegreen}{rgb}{0,0.6,0}
\definecolor{codegray}{rgb}{0.5,0.5,0.5}
\definecolor{codepurple}{rgb}{0.58,0,0.82}
\definecolor{backcolour}{rgb}{0.95,0.95,0.92}
\lstdefinestyle{mystyle}{
    backgroundcolor=\color{backcolour},   
    commentstyle=\color{codegreen},
    keywordstyle=\color{magenta},
    stringstyle=\color{codepurple},
    basicstyle=\ttfamily\footnotesize,
    breakatwhitespace=true,         
    breaklines=true,                 
    captionpos=b,                    
    keepspaces=true,             
    showspaces=false,                
    showstringspaces=false,
    showtabs=false,                  
    tabsize=2
}

\title{Nonlinear EM-based Signal Processing

\thanks{This work was partially supported by the European Union under the Italian National Recovery and Resilience Plan (NRRP) of NextGenerationEU, partnership on “Telecommunications of the Future” (PE00000001 - program “RESTART”), Structural Project S12 - Spoke 7, and by the EU Horizon project TIMES (Grant no. 101096307). Giulia Torcolacci was funded by an NRRP Ph.D. grant.}
}

\author{\IEEEauthorblockN{
Mattia Fabiani\IEEEauthorrefmark{1}\IEEEauthorrefmark{2},
Giulia Torcolacci\IEEEauthorrefmark{1}\IEEEauthorrefmark{2},
Davide Dardari\IEEEauthorrefmark{1}\IEEEauthorrefmark{2},
}\\
\vspace{3mm}
\IEEEauthorblockA{\IEEEauthorrefmark{1} DEI, University of Bologna, 40136 Bologna, Italy}\\
\IEEEauthorblockA{\IEEEauthorrefmark{2} National Laboratory of Wireless Communications (WiLab), CNIT, 40136 Bologna, Italy}
}

\maketitle

\begin{abstract}
The use of high-frequency bands, combined with antenna arrays containing an extremely large number of elements (XL-MIMO), is pushing current technology to its limits in terms of hardware complexity, latency, and power consumption. A promising approach to achieving scalable and sustainable solutions is to shift part of the signal processing directly into the electromagnetic (EM) domain. In this paper, we investigate novel architectures that harness the interaction of reconfigurable passive linear and nonlinear (NL) scattering elements positioned in the reactive near field of signal sources. The objective is to enable multifunctional linear and NL EM signal processing to occur directly “over-the-air.” Numerical results highlight the potential to significantly reduce both system complexity and the number of RF chains, while still achieving key performance metrics in applications such as direction-of-arrival and position estimation, without the need for additional analog or digital processing.
\end{abstract}

\providecommand{\keywords}[1]
{
  \textbf{\mathrm{Keywords: }} #1
}

\begin{IEEEkeywords}
NL-SIM, nonlinear, near-field, localization
\end{IEEEkeywords}



\acrodef{AI}{artificial intelligence}
\acrodef{AOA}{angle-of-arrival}
\acrodef{AOD}{angle-of-departure}
\acrodef{CSI}{channel state information}
\acrodef{CNN}{convolutional neural network}
\acrodef{DL}{deep learning}
\acrodef{ML}{maximum likelihood}
\acrodef{mmWave}{millimeter wave}
\acrodef{NL}{nonlinear}
\acrodef{NL-SIM}{nonlinear SIM}
\acrodef{LoS}{line-of-sight}
\acrodef{NLoS}{non-line-of-sight}
\acrodef{NN}{neural network}
\acrodef{OTA}{over-the-air}
\acrodef{RF}{radio-frequency}
\acrodef{RMSE}{root mean square error}
\acrodef{SIM}{stacked intelligent metasurface}
\acrodef{THz}{terahertz}
\acrodef{ULA}{uniform linear array}
\acrodef{UPA}{uniform planar array}
\acrodef{mMIMO}{massive multi-input multi-output}
\acrodef{ESP}{electromagnetic signal processing}
\acrodef{DSA}{dynamic scattering array}
\acrodef{DMA}{dynamic metasurface antenna}
\acrodef{SCM}{self-conjugating metasurface}
\acrodef{EM}{electromagnetic}
\acrodef{DOF}{degrees of freedom}
\acrodef{6G}{6th generation}
\acrodef{DNN}{deep neural network}
\acrodef{NN}{neural network}
\acrodef{ADC}{analog-to-digital converter}
\acrodef{AWGN}{additive white Gaussian noise}
\acrodef{UE}{user equipment}

\newcommand{\SNR}{\text{SNR}}
\newcommand{\SINR}{\text{SINR}}
\newcommand{\TNR}{\mathsf{TNR}}
\newcommand{\sigmaN} {\sigma_{\text{N}}}
\newcommand{\MSE} {\text{MSE}}

\newcommand{\bolda}{\mathbf{a}}
\newcommand{\boldb}{\mathbf{b}}
\newcommand{\boldbeta}{{\boldsymbol{\beta}}}
\newcommand{\boldc}{\mathbf{c}}
\newcommand{\boldd}{\mathbf{d}}
\newcommand{\bolde}{\mathbf{e}}
\newcommand{\boldf}{\mathbf{f}}
\newcommand{\boldg}{\mathbf{g}}
\newcommand{\boldh}{\mathbf{h}}
\newcommand{\boldi}{\mathbf{i}}
\newcommand{\boldj}{\mathbf{j}}
\newcommand{\boldk}{\mathbf{k}}
\newcommand{\boldl}{\mathbf{l}}
\newcommand{\boldm}{\mathbf{m}}
\newcommand{\boldn}{\mathbf{n}}
\newcommand{\boldo}{\mathbf{o}}
\newcommand{\boldp}{\mathbf{p}}
\newcommand{\boldq}{\mathbf{q}}
\newcommand{\boldr}{\mathbf{r}}
\newcommand{\bolds}{\mathbf{s}}
\newcommand{\boldt}{\mathbf{t}}
\newcommand{\boldu}{\mathbf{u}}
\newcommand{\boldv}{\mathbf{v}}
\newcommand{\boldw}{\mathbf{w}}
\newcommand{\boldx}{\mathbf{x}}
\newcommand{\boldy}{\mathbf{y}}
\newcommand{\boldz}{\mathbf{z}}
\newcommand{\Pt}{P_{\text{T}}}
\newcommand{\Nr}{N_{\text{R}}}

\newcommand{\boldA}{\mathbf{A}}
\newcommand{\boldB}{\mathbf{B}}
\newcommand{\boldC}{\mathbf{C}}
\newcommand{\boldD}{\mathbf{D}}
\newcommand{\boldE}{\mathbf{E}}
\newcommand{\boldF}{\mathbf{F}}
\newcommand{\boldG}{\mathbf{G}}
\newcommand{\boldH}{\mathbf{H}}
\newcommand{\boldI}{\mathbf{I}}
\newcommand{\boldJ}{\mathbf{J}}
\newcommand{\boldK}{\mathbf{K}}
\newcommand{\boldL}{\mathbf{L}}
\newcommand{\boldM}{\mathbf{M}}
\newcommand{\boldN}{\mathbf{N}}
\newcommand{\boldO}{\mathbf{O}}
\newcommand{\boldP}{\mathbf{P}}
\newcommand{\boldQ}{\mathbf{Q}}
\newcommand{\boldR}{\mathbf{R}}
\newcommand{\boldS}{\mathbf{S}}
\newcommand{\boldT}{\mathbf{T}}
\newcommand{\boldU}{\mathbf{U}}
\newcommand{\boldV}{\mathbf{V}}
\newcommand{\boldW}{\mathbf{W}}
\newcommand{\boldX}{\mathbf{X}}
\newcommand{\boldY}{\mathbf{Y}}
\newcommand{\boldZ}{\mathbf{Z}}

\section{Introduction}

\IEEEPARstart{F}{uture} \ac{6G} wireless networks, expected to support extreme data rates, ultra-low latency, and high reliability, will require a fundamental paradigm shift in physical-layer design \cite{ITU-M2160-framework}.
Meeting these requirements calls for higher carrier frequencies and electrically large antenna apertures capable of fine \ac{EM}-field control. The resulting increase in antenna elements and bandwidth naturally drives operation in the radiative near-field, where spherical wavefronts must be explicitly taken into account. At the same time, the scale of these arrays pushes the analog-digital interface closer to the antenna to limit losses and noise, demanding high-rate \acp{ADC} and numerous \ac{RF} chains. This configuration imposes a substantial digital burden, commonly referred to as the ``\textit{digital bottleneck}''.

Within this context, the emerging paradigm of \ac{ESP} has attracted significant attention \cite{ESP-concept}. \ac{ESP} proposes manipulating wireless signals directly at the \ac{EM} level, i.e., \textit{over-the-air}, using reconfigurable scattering structures operating passively to shape the \ac{EM} field before digital conversion, thus exploiting the full \ac{DOF} offered by the wireless channel. Notably, these over-the-air transformations operate at the speed of light with minimal active circuitry, enabling low-latency, energy-efficient implementations of key signal processing functions such as beamforming and spatial multiplexing \cite{Dar:J25_multifunc}.

Several architectures have been investigated as promising candidates for practical \ac{ESP} implementations.
Among them, \acp{SIM} emerged as a hardware platform to enable \ac{ESP}-based transceivers. A \ac{SIM} consists of multiple stacked metasurface layers enclosed in a vacuum box, each comprising reconfigurable cells applying tunable phase shifts to the incident \ac{EM} wavefront \cite{an2023stacked_holographic}. By operating according to the Huygens-Fresnel principle, each cell acts as a secondary radiating source, illuminating downstream layers and enabling a sequence of finely controllable interactions with the propagating \ac{EM} field. This multilayer interplay provides continuous wave-domain processing, allowing \acp{SIM} to approximate a wide class of linear transformations with a fraction of the hardware complexity required by conventional antenna arrays. Although such linear operators are adequate for some wave-based tasks like beam steering and mode conversion, they inherently lack the expressive power required by parameter estimators and data-driven functionalities that depend on \ac{NL} mappings, e.g., physical-domain classifiers or analog neural operators (see, e.g., \cite{gu2024classification}). Prior attempts at \ac{RF}-domain neural processing include \cite{gao2023programmable}, where a \ac{NN} is emulated using cascaded \ac{RF} components; however, \ac{NL} activations are implemented through an active feedback loop that requires continuous amplitude monitoring, limiting scalability and fully passive operation. Recent efforts toward \ac{NL} metasurface-based processing include the \ac{NL-SIM} architecture in \cite{abbas2025nonlinear}, which introduces nonlinearity through a tunable step-type suppression mechanism. However, this approach is constrained to a specific thresholding mechanism, limiting its applicability to more general cell-level \ac{NL}. In parallel, \cite{stylianopoulosUniversalApproximationXL2025} proposes distributing neural computation across the wireless channel, but the \ac{NL} activations are active, and the final combining is done in the digital domain, thus requiring a large number of \ac{RF} chains.

Motivated by these limitations, we introduce a \ac{NL} \ac{SIM} architecture, termed \ac{NL-SIM}, in which selected metasurface layers incorporate passive \ac{NL} cells that apply pointwise \ac{NL} transformations to the received \ac{EM} field directly at \ac{RF}. This configuration realizes an \ac{EM}-domain analog of a \ac{NN}, where conventional linear \ac{SIM} layers exploit the \ac{EM} superposition principle to realize linear transformations, while the \ac{NL} cells act as activation functions, thereby enhancing the representational power of the structure for complex tasks such as near-field sensing and localization.

Building on this concept, we develop a general end-to-end \ac{EM} transformation framework that rigorously captures the impact of element-wise \ac{NL} and clarifies the constraints they impose on the equivalent low-pass model commonly used in the analysis of wireless systems. We further propose a practical circuit-level realization of the passive \ac{NL} cell through a simple diode-based design, demonstrating its feasibility and physical consistency. Finally, by introducing a representative near-field localization use case, we show that \ac{NL-SIM} achieves comparable performance whether the \ac{NL} cell is fully trainable or controlled via fixed random biasing, offering clear advantages in terms of hardware simplicity and power consumption.

\section{NL-SIM for Nonlinear Wave-based Processing}
\label{sec:NL-SIM}

The core premise of the proposed \ac{NL-SIM} is to embed passive \ac{NL} elements within the cells of one or more metasurface layers, thereby leveraging their intrinsic input-output \ac{NL} relation to enrich the expressive power of the \ac{EM}-level signals transformation and enhance the overall end-to-end learning capability. In this context, the \ac{NL-SIM} can be conceptually viewed as an \ac{EM} analog of a \ac{NN}, but constrained by the physical laws governing waves propagation and cells' functioning. Leveraging this structural parallelism, the following analysis adopts standard terminology from the deep learning domain, including \textit{training}, \textit{bias}, and \textit{inference}, to describe, respectively, the configuration, optimization, and operation of the physical device.

\begin{figure}[t]
    \centering
    \includegraphics[width=.48\textwidth]{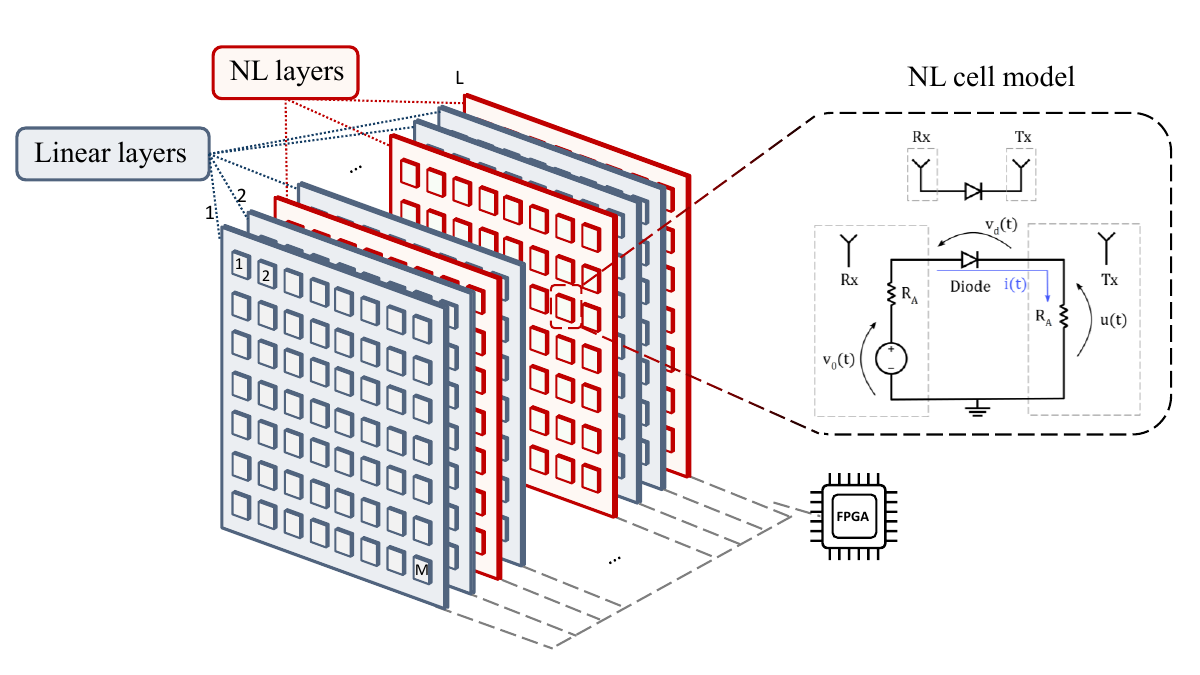}
    \caption{General NL-SIM architecture for wave-based signal processing.}
    \label{fig:SIMarchitecture}
\end{figure}

With reference to Fig.~\ref{fig:SIMarchitecture}, the considered \ac{NL-SIM} consists of a total number $L$ of metasurface layers, where selected layers implement \ac{NL} transformations and the remaining ones apply linear transformations.
Denote by $\mathcal{L}=\{1,2, \ldots, L\}$ the set of indices corresponding to all the metasurface layers and by $\mathcal{Q}\subset \mathcal{L} $ the subset of cardinality $Q$ of indices associated with the \ac{NL} metasurface layers.    
Each layer comprises a square grid of $M$ cells spaced by $\lambda/ 2$, with $\lambda = c/f_0$ denoting the wavelength at the carrier frequency $f_0$, and $c$ is the speed of light. In the case of linear metasurface layers, the transmission coefficient of the $m$th cells on the $l$th metasurface layer is represented by $\phi_m^{(l)} = e^{\jmath \theta_m^{(l)}}$, with $\theta_m^{(l)} \in [0,2\pi)$, for $m = 1, 2,  \dots, M$ and $l \in \mathcal{L}$, and $\jmath$ denoting the imaginary unit. 
Therefore, the matrix of transmission coefficients for the $l$th metasurface layer can be expressed as $\boldsymbol{\Phi}^{(l)} = \mathrm{diag}\left({ \phi_1^{(l)}, \phi_2^{(l)}, \ldots, \phi_M^{(l)}}\right) \in \mathbb{C}^{M \times M}$. 

Let us consider a generic equivalent complex low-pass signal $\bolds\in\mathbb{C}^{M\times1}$ associated with that impinging on the first metasurface layer of the \ac{SIM}. 
The signal $\boldx^{(L)}\in\mathbb{C}^{M\times1}$ at the output of the \ac{NL-SIM} architecture can be obtained recursively as 
\begin{equation}
    \begin{cases}
    \boldx^{(0)} = \bolds, \\[4pt]
    \boldx^{(i)} = 
    \begin{cases}
        \boldsymbol{\sigma}^{(i)}\!\left(\boldW^{(i)} \boldx^{(i-1)}\right)\,, & i\in\mathcal{Q} \\
        \boldsymbol{\Phi}^{(i)} \boldW^{(i)} \boldx^{(i-1)}\,, & \text{otherwise}
    \end{cases}
    \end{cases}
    \label{eq:NL_transferfunc}
\end{equation}
with $i \in \mathcal{L}$ and $\boldW^{(1)} = \boldI_M$. Here, the matrices $\boldW^{(l)} \in \mathbb{C}^{M \times M}$, $l \in \mathcal{L}$, model the propagation between the $(l-1)$th and $l$th metasurface layer according to the Rayleigh-Sommerfeld diffraction theory~\cite{an2023stacked_holographic}, whose $(m,i)$th entry is given by
\begin{equation}\label{eq:intralayprop}
\left[\boldW^{(l)}\right]_{m,i} = 
\frac{A \cos\chi_{m,i}^{(l)}}{d_{m,i}^{(l)}} 
\left(\frac{1}{2\pi d_{m,i}^{(l)}} - \frac{\jmath}{\lambda}\right)
e^{\jmath k d_{m,i}^{(l)}} \, ,
\end{equation}
with $k = 2\pi/\lambda$ being the wavenumber, $d_{m,i}^{(l)}$ representing the propagation distance between the $i$th cell of the $(l-1)$th layer and the $m$th cell of the $l$th layer, $\chi_{m,i}^{(l)}$ denotes the angle between the propagation direction and the cell's normal, while $A=\lambda^2/4$ is the area of each cell.

In addition, $\boldsymbol{\sigma}^{(i)} = \left[\sigma_1^{(i)}(\cdot),\, \dots, \sigma_m^{(i)}(\cdot),\, \dots,\sigma_M^{(i)}(\cdot)\right]\in\mathbb{C}^{M\times1}$, $i \in \mathcal{Q}$, represents the vector of \ac{NL} responses applied by each cell of the $i$th \ac{NL} layer to the impinging \ac{EM} signal. In general, the set of admissible \ac{NL} activation functions cannot be selected arbitrarily. First, they must comply with the standard conditions required by the universal approximation theorem \cite{Les:93} (e.g., the activation must not degenerate into a purely polynomial function). Second, as the processing is performed over the air (i.e., at \ac{RF}), the effective \ac{NL} response in the low-pass equivalent representation deviates from that obtained at \ac{RF}; hence, the physically implemented \ac{NL} activation may induce a different complex low-pass behavior, as further detailed in the following section.

\section{Nonlinearities Modelling at RF}
\label{sec: nonlinearities}

In this section, we first clarify the link between the RF (i.e., bandpass) \ac{NL} response and its low-pass equivalent representation. We then introduce two diode-based implementations of the \ac{NL} cell. The resulting \ac{NL} behavior can be either trainable or fixed: in the trainable case, the diode structure is unchanged and its operating point is optimized during learning, whereas in the fixed case the operating point is predetermined and diversity is introduced by randomly sampling the diode doping parameter.

\subsection{Nonlinearity: Bandpass vs. Low-pass}
Consider a generically modulated narrowband \ac{RF} signal $s(t) = V(t)\cos(2\pi f_0 t + \phi(t))$, with $V(t)\geq 0$, and its equivalent complex low-pass representation $x(t) = V(t)\, e^{j \phi(t)}$. Suppose the signal passes through a memoryless \ac{NL} device exhibiting a bandpass response $u(t) = F[s(t)]$, followed by a zonal filter eliminating all the harmonic components except the fundamental at $f_0$. Denote with $y(t)$ the complex low-pass version of the resulting filtered output $\tilde{u}(t)$. The following theorem holds

\begin{theorem}[]
The \ac{NL} relationship between $y(t)$ and $x(t)$ is given by 
\begin{equation} \label{eq:sigma}
y(t) = \sigma(x(t))=C\left[|x(t)|\right] \, e^{\jmath \arg\{ x(t) \}} \, ,
\end{equation}
where the \ac{NL} function $C[v]$ is given by
\begin{equation} \label{eq:C}
C[v]=\frac{2}{\pi} \int_{0}^{\pi} F[v\, \cos(\phi)] \, \cos(\phi) \, \text{d}\phi \, .
\end{equation}
\end{theorem}

From \eqref{eq:sigma}, it follows that the equivalent low-pass response is governed by a \ac{NL} relationship, which is in general different from that at \ac{RF} $F[v]$. Moreover, the following properties hold: (\textit{i}) only the amplitude of the input signal is affected by the \ac{NL}, whereas the argument is unchanged; (\textit{ii}) if $F[v]$ is an even function, then $C[v]=0$ and the output is zero.

In the following, we report explicit expressions for $C[v]$ for some common \ac{NL} functions $F[v]$. Specifically:

\begin{itemize}
    \item $F[v] = \max(v, 0)$ (ReLU): $C[v] = v/2$ (\emph{Linear response})
    \item $F[v] = \max(v+a, 0)$ \emph{(Shifted ReLU)}: \\[5pt]$C[v]=\left \{  
\begin{array}{cc}
\displaystyle \frac{1+\operatorname{sgn}a}{2}\,v, 
& v<|a|,\\[6pt]
\displaystyle \frac{1}{\pi}\!\left(v\arccos\!\!\left(-\frac{a}{v}\right)
+\frac{a\sqrt{v^{2}-a^{2}}}{v}\right), 
& v>|a|.
\end{array}
 \right .$\\[5pt]
    \item $F[v] = |v|$: $C[v] = 0$ $\Rightarrow$ No output (even function).
    \item $F[v] = \text{sgn}(v)$: $C[V] = \frac{4}{\pi}$ $\Rightarrow$ Constant amplitude output.
    \item $F[v] = v^n$: $C[v] = \frac{\Gamma \left (1 + \frac{n}{2} \right )}{\sqrt{\pi} \Gamma \left ( \frac{3 + n}{2} \right )}=\frac{1}{2^n} \binom{n}{\frac{n+1}{2}}$, for $n$ odd and zero otherwise, where $\Gamma[x]$ denotes the Eulero gamma function \cite{ConDarTra:C03}. 
\end{itemize}

\subsection{Example of RF Nonlinear Activation}

\begin{figure}[t]
    \centering
    \includegraphics[width=.48\textwidth]{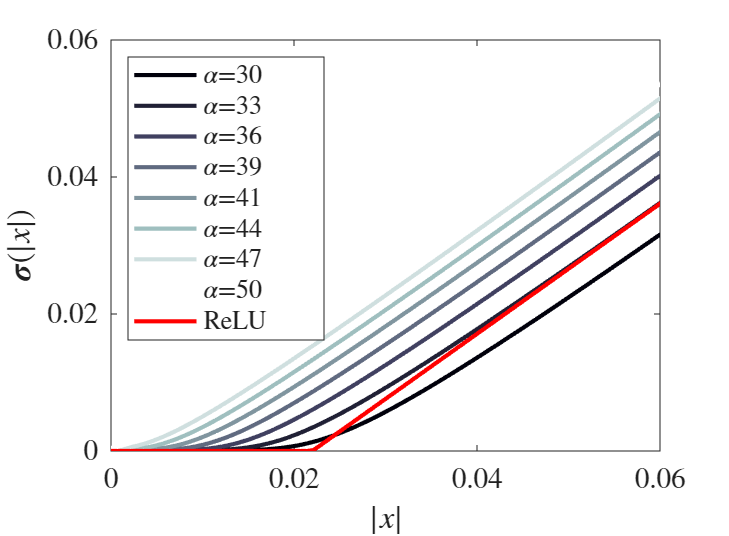}
    \caption{Examples of NL activation functions having a common bias voltage of $b=0.4\,\mathrm{V}$ for varying $\alpha$ values. The ReLU approximation for the $\alpha=33$ case is reported in red.}
    \label{fig: NL curves}
\end{figure}
As a simple illustrative example, consider a practical \ac{NL} activation cell in which a receive antenna is connected to a transmit antenna on the opposite side of the surface via a diode, as shown in the inset of Fig.~\ref{fig:SIMarchitecture}. In the same figure, the corresponding equivalent circuit for resonant antennas at frequency $f_0$ (e.g., half-wave dipoles) is also reported, where $R_{\text{A}}$ denotes the antenna radiation resistance, $v_0(t)$ the open-circuit voltage of the receiving antenna induced by the impinging signal, $i(t)$ the current flowing in the circuit, and $v_d(t)$ the voltage across the diode.\footnote{The reactive elements of the equivalent circuit are omitted, as they are assumed to be counterbalanced through complex-conjugate impedance matching.}

The \ac{NL} current-voltage characteristic of a diode is given by
\begin{equation}\label{eq:diode}
i(t) = I_s \left( e^{\alpha v_d(t)} - 1 \right)\,,
\end{equation}
where $I_s$ is the diode's saturation current and $\alpha$ depends on the thermal voltage and diode fabrication characteristics. The zonal filter is omitted, as the antenna itself acts as a filter for higher harmonics.

Denoting by $P_a$ the available power of $s(t)$ (and thus $x(t)$) at the input antenna, i.e., the power that would be delivered to a perfectly matched load, it follows that $P_a=\mathbb{E}\{v_0(t)\}^2/(4 R_{\text{A}})$ and $v_0(t)=2 s(t)$, where $\mathbb{E}\{ \cdot \}$ represents the statistical expectation.
The (unfiltered) transmitted signal corresponds to the voltage across the transmit antenna, i.e., $u(t)=R_{\text{A}} i(t)$. Therefore, the \ac{NL} input-output relationship $F[\cdot]$ can be obtained by analyzing the circuit in Fig. \ref{fig:SIMarchitecture}, which is governed by the following transcendental equation
\begin{equation}\label{eq:NLcell}
u(t) = R_{\text{A}} I_s \left( e^{\alpha 2 (s(t) -  u(t)))} - 1 \right) \, .
\end{equation}

The solution of \eqref{eq:NLcell} and \eqref{eq:C} can be evaluated numerically and possibly approximated with an analytically tractable expression, thus obtaining the cell's activation function $\sigma(x)$ in \eqref{eq:sigma}. In order to enhance the approximation capability of the NL-SIM, the activation functions must vary cell-by-cell. To this purpose, we consider two implementation paradigms: 

\subsubsection{Trainable NL cells}
In the trainable configuration, the diode-based $m$th cell of the $i$th layer is fixed, but its operating point is optimized during training by adding a bias $b_m^{(i)} \le 0$ to the input signal, which shifts the diode's current-voltage curve, hence modifying the shape of the induced \ac{NL} activation. In this case, $\sigma_m^{(i)}(x)$ in \eqref{eq:NL_transferfunc} is given by $\sigma_m^{(i)}(x)=\sigma\left (x+b_m^{(i)} \right )$.  During the learning phase, each cell in the \ac{NL} layer(s) adapts its bias so that the resulting \ac{NL} extracts relevant features from the incoming signal. This can be physically implemented by embedding a controllable bias network within each cell or group of cells.

\subsubsection{Static random NL cells}
In the non-trainable case, the operating point is fixed for all elements, and no bias tuning occurs during learning. Instead, diversity is introduced at fabrication time by sampling the diode parameters such as $\alpha$ (e.g., doping concentration or junction area) from a predefined distribution. This yields a set of static \ac{NL} activation curves $\sigma_m^{(i)}(x)=\sigma_{\alpha_{m,i}}(x)$, each slightly shifted horizontally due to manufacturing variations, where $\sigma_{\alpha_{m,i}}$ is obtained by solving \eqref{eq:NLcell} with parameter $\alpha=\alpha_{m,i}$ randomly chosen within a certain interval $[\alpha_{\text{min}},\alpha_{\text{max}}]$. This approach is particularly appealing for fully passive deployments, as it avoids active biasing networks and reduces implementation costs while still enabling \ac{NL} wave processing, as demonstrated in the numerical study.

In this regard, Fig.~\ref{fig: NL curves} shows several \ac{NL} activation functions corresponding to different $\alpha$ values. As illustrated, varying $\alpha$ results in distinct \ac{NL} responses. For numerical efficiency, each curve is approximated using a ReLU function in the simulations, without significantly affecting the overall behavior.

\section{Use Case: Near-field Localization}
\label{sec: numerical results}

\begin{figure}[t]
    \centering
    \includegraphics[width=.48\textwidth]{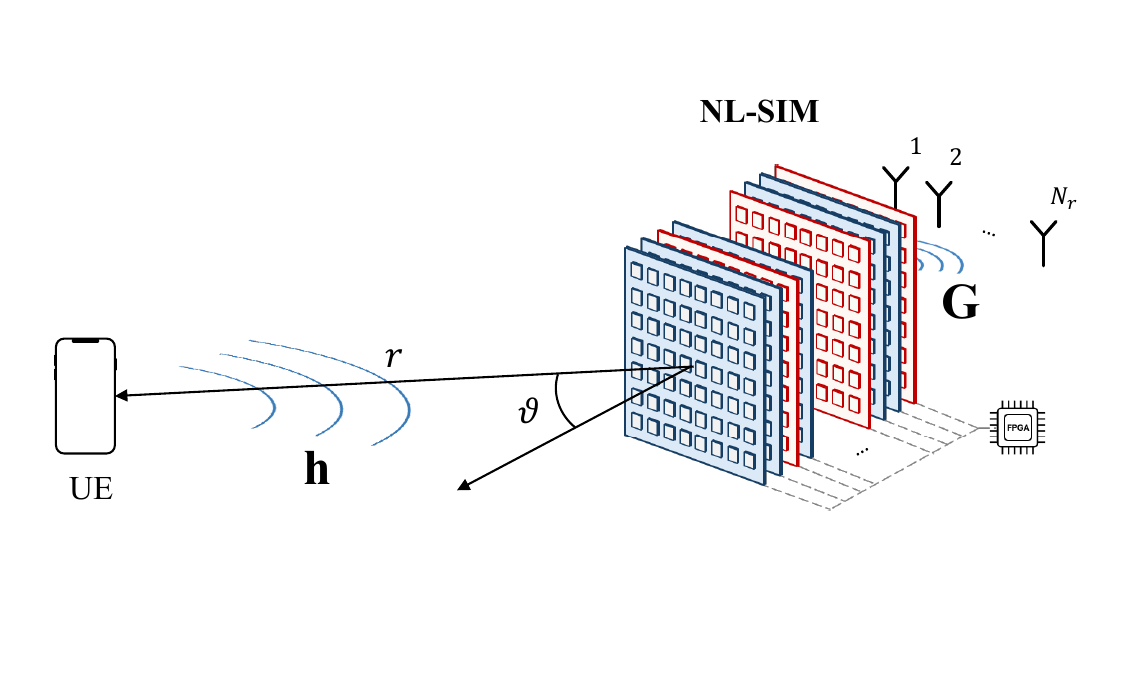}
    \caption{Considered scenario incorporating the proposed \ac{NL-SIM} architecture for localization tasks in the near-field.}
    \label{fig:scenario}
\end{figure}

This section illustrates a representative application of the proposed \ac{NL-SIM} architecture for near-field localization. 
Let us consider the scenario depicted in  Fig.~\ref{fig:scenario}, where a single-antenna \ac{UE} transmits a narrowband pilot symbol $s \in\mathbb{C}$ toward the \ac{NL-SIM}-based receiver. The receiving architecture comprises a \ac{NL-SIM} followed by an \ac{ULA} with $\Nr$ antenna elements, which is located at a distance $d$ from the last \ac{NL-SIM} layer. The \ac{UE} is located in the radiative near-field region of the receiver, specifically at a distance $r \approx 0.1 d_{\mathrm{F}}$ from the receiver, being $d_{\mathrm{F}} = 2D^2/\lambda$ the Fraunhofer distance \cite{BalB:16}.
The goal is to estimate the \ac{UE} position instantaneously from its uplink pilot by leveraging the intrinsic wavefront curvature in the near-field, using a minimal number of \ac{RF} chains and offloading part of the conventional baseband processing to the \ac{NL-SIM}.

We define a Cartesian coordinate system centered in the first \ac{NL-SIM} layer center. The \ac{UE} position is given by the polar coordinates $(r, \theta)$, where $r$ denotes the radial distance from the \ac{NL-SIM} aperture center, and $\theta$ represents the azimuth angle relative to the \ac{SIM} aperture boresight (broadside direction). The near-field array response vector $\mathbf{a}(r, \theta) \in \mathbb{C}^{M\times1}$ is expressed as 
\begin{equation}
    \mathbf{a}(r, \theta) = \frac{1}{\sqrt{M}} 
    \left[1, \dots, e^{-\jmath k(r - r_m)}, \dots, e^{-\jmath k(r - r_{M-1})}\right]^{\mathsf{T}},
\end{equation}
where $r_m$ represents the Euclidean distance between the $m$th cell of the first layer and the \ac{UE}. Furthermore, the near-field channel $\mathbf{h} \in \mathbb{C}^{M\times1}$  between the \ac{UE} and the \ac{SIM} input layer is modeled as a Rician fading channel vector according to
\begin{equation}
    \mathbf{h} = 
    \frac{1}{\sqrt{P_L}}
    \left(
    \sqrt{\frac{\kappa}{\kappa + 1}}\, \mathbf{a}(r, \theta)\,e^{\jmath \gamma}
    + \sqrt{\frac{1}{\kappa + 1}}\, \mathbf{h}_{\mathrm{NLoS}}
    \right),
\end{equation}
where $P_L$ denotes the path loss, $\kappa$ is the Rician factor, and $\gamma \sim \mathcal{U}[0, 2\pi)$ denotes a random initial phase. Moreover, $\mathbf{h}_{\mathrm{NLoS}}$ represents the \ac{NLoS} component, capturing scattering effects, and is modelled as a Rayleigh fading vector with entries independently drawn from $\mathcal{CN}\left(0, \frac{1}{M}\right)$.

To perform wave-domain localization, the \ac{NL-SIM} processes the incident \ac{EM} field layer-by-layer. Let us define the signal impinging on the input \ac{NL-SIM} layer as 
\begin{equation}
    \mathbf{s} = \mathbf{h}s + \mathbf{n}\, ,
    \label{eq:input_layer}
\end{equation}
where $\mathbf{n}\in\mathbb{C}^{M\times 1}$, $\boldn \sim\mathcal{CN}(0, \sigma_{\mathrm{N}}^2 )$, is the complex \ac{AWGN} vector. Such signal then propagates through the $L$ \ac{NL-SIM} layers according to \eqref{eq:NL_transferfunc}. Therefore, the signal $\mathbf{y}\in\mathbb{C}^{\Nr\times 1}$ that is received at the \ac{ULA} is given by
\begin{equation}
    \mathbf{y} = \mathbf{G}\mathbf{x}^{(L)},
\end{equation}
where $\mathbf{G}\in\mathbb{C}^{\Nr\times M}$ denotes the channel between the last \ac{NL-SIM} layer and the receiving \ac{ULA}, whose elements can be obtained by replacing $d_{m, i}^{(l)}$ in \eqref{eq:intralayprop} with the distance between the each \ac{ULA} element and the $m$th cell of the last \ac{NL-SIM} layer.

Notably, the \ac{NL-SIM} is engineered to extract the dominant near-field features, allowing the \ac{ULA} to operate with simple amplitude-only measurements. The \ac{NL-SIM} is thus trained to perform a physical regression task, mapping the received signal magnitudes directly to the \ac{UE} position parameters $(r,\theta)$. Specifically, for $\Nr=2$, this mapping reduces to
\begin{equation}
    \begin{cases}
        \hat{r} = r_\mathrm{min} + \beta \left|y_1\right| (r_\mathrm{max} - r_\mathrm{min}) \, , \\[8pt]
        \hat{\theta} = (2\beta |y_2|-1) \frac{\pi}{2}\, ,
    \end{cases}
    \label{eq:y_to_r_theta}
\end{equation}
where $[r_\mathrm{min}, r_\mathrm{max}]$ denotes the spatial range of interest, and $\beta$ is a constant scaling factor such that $\beta \, \mathrm{max}(\left|y_i\right|)\approx1\,, i\in\{1,2\}$. The \ac{NL-SIM} is explicitly trained to focus the signal onto each output antenna so that the measured amplitudes $|y_1|$ and $|y_2|$ provide a direct, linear indication of the corresponding target's range and angle parameters, respectively. Finally, the estimated \ac{UE} position is recovered as $\hat{\mathbf{p}}=[\hat{r}\cos\hat{\theta}, \hat{r}\sin\hat{\theta}]$, which is the only operation performed in the digital domain.

\section{Numerical Results}
\label{sec:results}

We hereby assess the performance of \ac{NL-SIM}-based processing in terms of positioning accuracy, comparing it to a standard linear \ac{SIM} to highlight the gains of \ac{NL} wave-based operations. Comparisons with fully-digital schemes further illustrate that \ac{NL-SIM} achieves competitive accuracy with substantially lower hardware complexity.

\subsection{Simulation Setup}

We generate a synthetic dataset of $N=10^4$ channel realizations, assuming a carrier frequency of $f_0 = 28$~GHz. The \ac{UE} positions are sampled uniformly in the near-field region, with radial distances $r \in [1, 3]$~m and azimuth angles $\theta \in [-70^{\circ}, 70^{\circ}]$, ensuring an even spatial coverage. The channel follows a Rician fading model with factor $\kappa = 20$~dB. The transmit power is set to $\Pt = 30$~dBm, while the receiver noise power is $\sigma_{\mathrm{N}}^2 = -110$~dBm.

The considered receiving architecture comprises a \ac{ULA} with $\Nr=2$ antenna elements and, unless otherwise stated, a \ac{NL-SIM} with a total number $L=6$ of metasurface layers. Each layer consists of $M=1600$ cells arranged as a $40\times40$ square grid. The distance between the last \ac{NL-SIM} layer and the \ac{ULA} is set to $d=3\lambda$. To stabilize training, i.e., bounding the magnitude of the backpropagated error gradients, the range and angle ground-truth are scaled to the range $[-1,1]$ via
\begin{equation}
    \begin{cases}
        r' = \dfrac{r - r_{\min}}{r_{\max} - r_{\min}} \cdot 2 - 1, \\[8pt]
        \theta' = \dfrac{2\,\theta}{\pi}.
    \end{cases}
    \label{eq: (r,theta) to [-1,1]}
\end{equation}
The dataset is split into training, validation, and test sets, comprising, respectively, $80\%$, $10\%$, and $10\%$ of the total dataset size $N$.

To update the \ac{NL-SIM} weights during the training process, we utilize the \ac{RMSE} of the estimated \ac{UE} position as the loss function. The \ac{RMSE} also serves as a performance metric to assess positioning accuracy. Specifically, it is defined as
\begin{equation}\label{eq:loss}
    \mathrm{RMSE} = \sqrt{\frac{1}{N_{\mathrm{test}}}\sum_{i=1}^{N_{\mathrm{test}}} \|\mathbf{p}_i - \hat{\mathbf{p}}_i\|^2 },
\end{equation}
where $N_{\mathrm{test}} = 0.1\, N$ is the test set size, $\mathbf{p}_i$ is the ground-truth position vector, and $\hat{\mathbf{p}}_i$ is the position estimate.

All learning models are implemented in PyTorch. We leverage PyTorch's native support for automatic differentiation to calculate via backpropagation the gradient of the loss function with respect to the physical parameters, namely, the phase shifts $\boldsymbol{\Phi}^{(i)}$ and the biases $b_m^{(i)}$. 
The \ac{SIM} is trained using the Adam optimizer, treating the physical wave propagation as a differentiable forward pass. For comparison, the following architectures are evaluated:

\subsubsection{\textbf{NL-SIM}}
We consider a \ac{NL-SIM} composed of $L$ layers, where only one layer is equipped with \ac{NL} activations, i.e., $Q=1$. 
Specifically, for the static random \ac{NL} cells, biases are first drawn from a standard Gaussian distribution, $b_m \sim \mathcal{N}(0,1)$, then rectified and rescaled as $b_m \leftarrow -|b_m|/10^5$, shifting the \ac{NL} curves to the right and producing a half-normal distribution with standard deviation $10^{-5}$. Under the diode model of Fig.~\ref{fig: NL curves}, this corresponds to the \ac{NL} coefficient $\alpha$ approximately in the range $[55, 57]$. For trainable \ac{NL} cells, $\alpha$ is initialized similarly but updated during training, while for static random cells the parameters remain fixed.

\subsubsection{\textbf{Linear SIM}} In this case, the conventional linear \ac{SIM} architecture with $L$ layers is considered, whose unit-cells apply exclusively linear phase shifts (i.e., $\mathcal{Q} = \emptyset$). This case allows isolating the performance gain specifically provided by the \ac{NL} activation functions.

\subsubsection{\textbf{Fully-Digital CNN}} A \ac{CNN} takes as input the raw baseband signal vector $\mathbf{s}$ in \eqref{eq:input_layer} received at the \ac{NL-SIM} side formatted as a $(\sqrt{M}\times\sqrt{M}\times 2)$ tensor, so that to embed both real and imaginary components. In particular, the \ac{CNN} architecture consists of three stacked convolutional blocks (namely, 2D Conv $3\times3$, ReLU, Dropout with probability $p=0.1$) followed first by a fully connected layer and then by a $\tanh$ activation function, which bounds the output to $[-1, 1]$; the estimated position is then obtained by inverting \eqref{eq: (r,theta) to [-1,1]}.

\subsubsection{\textbf{Maximum Likelihood (ML)}}
The \ac{ML} estimator is implemented as a grid-based exhaustive search, which serves as a performance lower bound. Given the received signal in \eqref{eq:input_layer} and the location-dependent array response vector $\mathbf{a}(r, \theta)$, the \ac{ML} estimate is obtained by maximizing the beamforming gain (spatial matched filtering) over a discretized domain $\mathcal{G}$ as 
\begin{equation}
    (\hat{r}, \hat{\theta})_{\mathrm{ML}} = \operatorname*{arg\,max}_{(r, \theta) \in \mathcal{G}} \left| \mathbf{a}^{\mathsf{H}}(r, \theta) \, \mathbf{y} \right|^2,
    \label{eq:ML_grid_search}
\end{equation}
where $(\cdot)^{\mathsf{H}}$ denotes the conjugate transpose. The grid $\mathcal{G}$ is defined with resolution $N_{\theta} \times N_{r}$, with $N_r=N_\theta=1000$, within the near-field region.

\subsection{Impact of Non-Linear Layer Placement}
\begin{figure}[t]
    \centering
    \includegraphics[width=.45\textwidth]{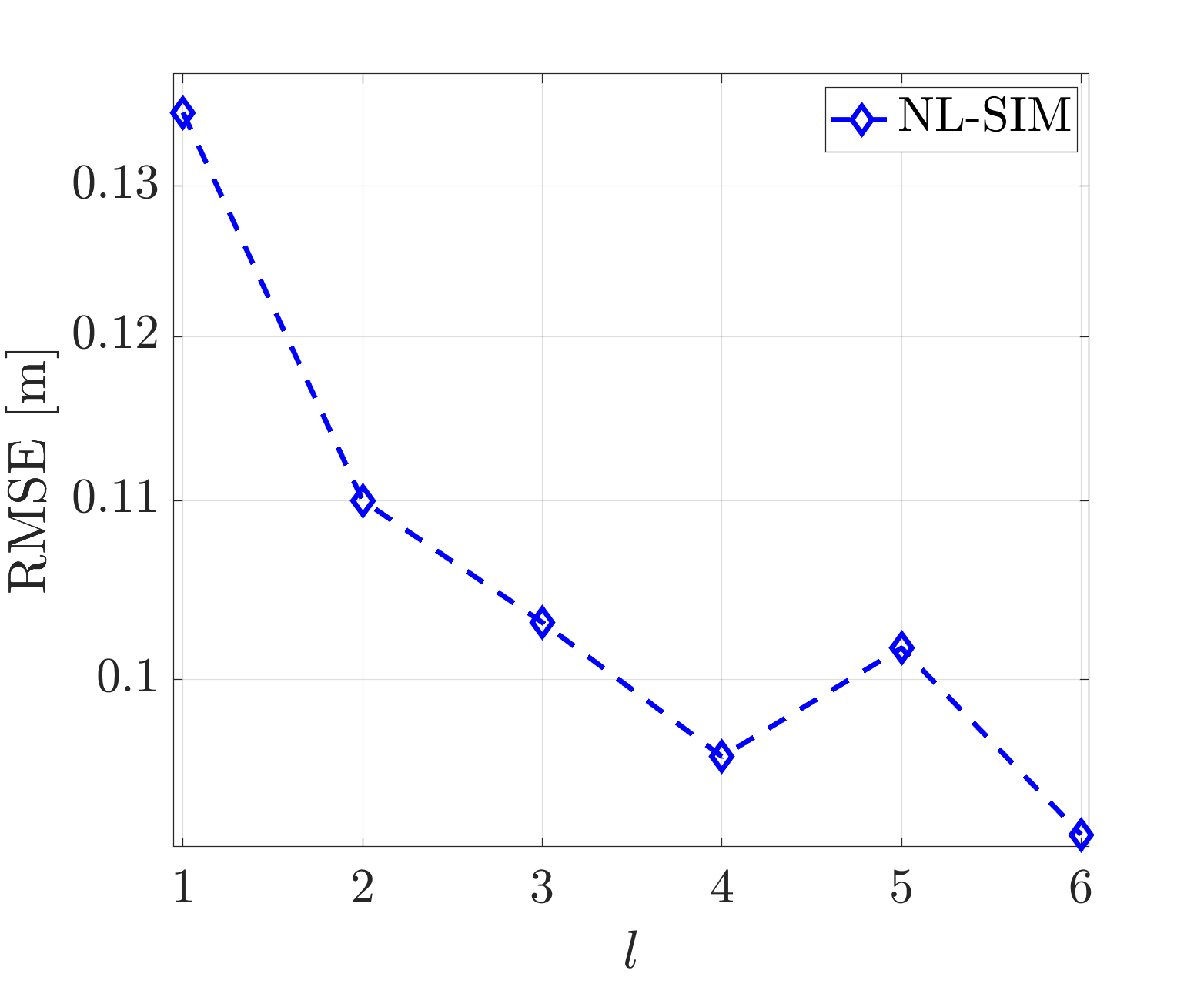}
    \caption{Position estimation \ac{RMSE} as a function of the \ac{NL} layer position $l$, $l \in \mathcal{L}$, for the case $L=6$ and $Q=1$.}
    \label{fig:results_NL_layer}
\end{figure}
First, we investigate the optimal \ac{NL-SIM} architecture by analyzing how the placement of the \ac{NL} activation layer influences localization accuracy. Specifically, we consider a \ac{NL-SIM} in which the index $l \in \mathcal{L}$ (i.e., the layer position) of the \ac{NL} layer is varied.

Fig.~\ref{fig:results_NL_layer} shows the position estimation \ac{RMSE} as a function of the \ac{NL} layer index in the \ac{NL-SIM} stack. The performance is highly sensitive to the placement of the \ac{NL} processing, with the minimum \ac{RMSE} achieved when the \ac{NL} layer is applied at the final stage of the stack ($\mathcal{Q}=\{L\}$). This observation aligns with standard \ac{NN} design, where \ac{NL} activations follow linear projection. Here, the preceding linear \ac{SIM} layers collectively implement a linear matrix. Positioning the \ac{NL} layer at the output enables it to act as a feature discriminator, shaping the final intensity distribution for accurate parameter readout. Conversely, placing \ac{NL} earlier exposes the generated features to subsequent linear mixing, which attenuates the near-field information essential for the amplitude-to-parameter mapping in \eqref{eq:y_to_r_theta}. Based on this, we set $\mathcal{Q}=\{{L}\}$ for the subsequent evaluation.

\subsection{Impact of Architecture Depth}

\begin{figure}[t]
    \centering
    \includegraphics[width=.48\textwidth]{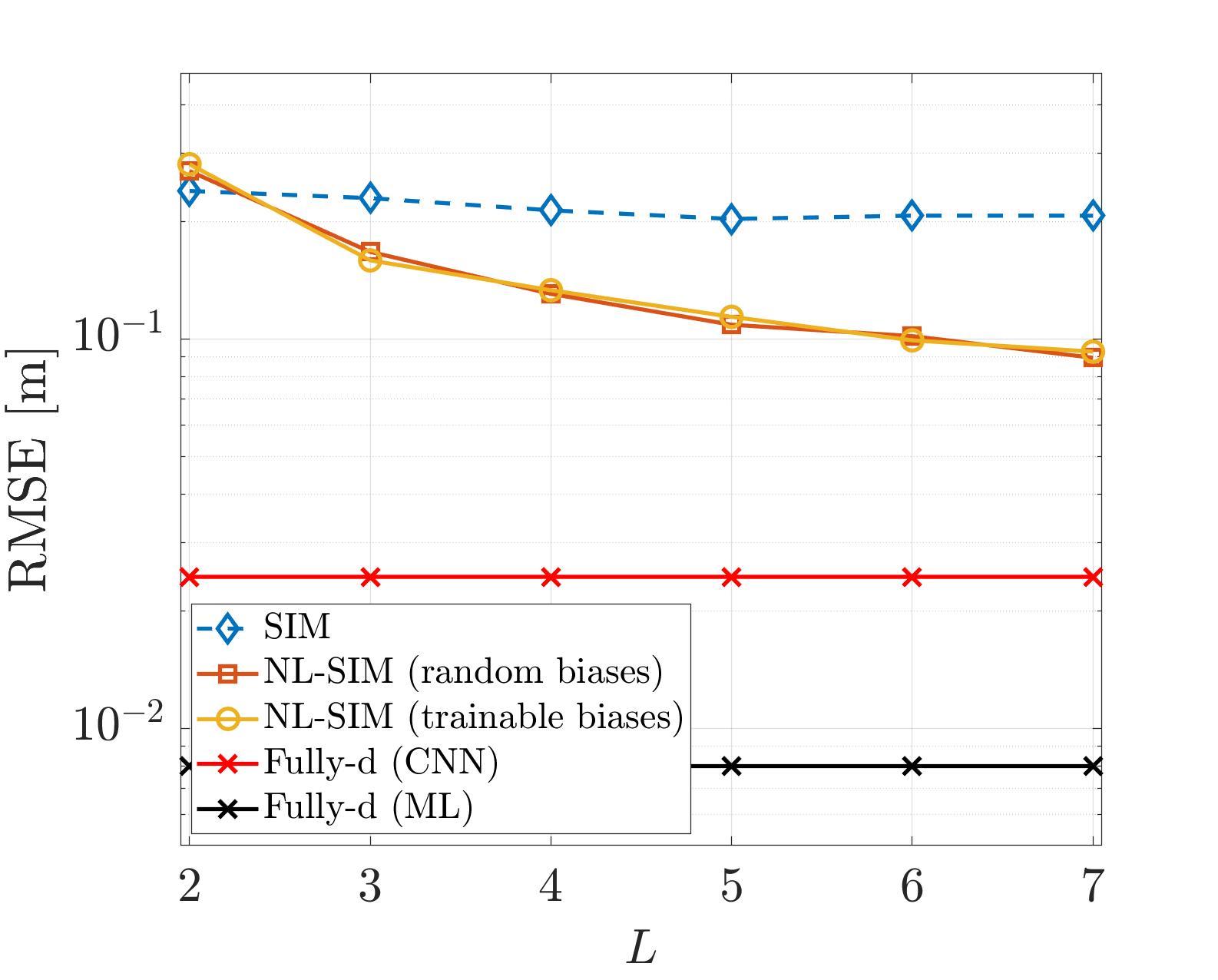}
    \caption{Position \ac{RMSE} as a function of the total number of \ac{SIM} layers $L$ for $\mathcal{Q}=\{L\}$.}
    \label{fig:results_L_layers}
\end{figure}

We further evaluate the scalability of the proposed approach by varying the total number of \ac{SIM} layers $L$. As shown in Fig.~\ref{fig:results_L_layers}, increasing the architecture depth leads to a reduction in the \ac{RMSE} for the \ac{NL-SIM} case, whereas the linear \ac{SIM} shows negligible improvement. Hence, a deeper \ac{SIM} architecture offers a larger number of trainable parameters and more complex wave processing capabilities.

Notably, the performance of the \ac{NL-SIM} with trainable biases is almost identical to the configuration with fixed random biases.  This is a crucial finding from a hardware implementation perspective: it implies that complex active biasing circuits are not strictly necessary. Instead, as discussed in Sec.~\ref{sec: nonlinearities}, we can rely on fully passive diodes with a fixed bias term, significantly reducing power consumption and circuitry complexity without compromising accuracy.

However, a performance gap remains between the \ac{NL-SIM} and the fully-digital systems. Compared to the \ac{SIM}, the \ac{CNN} achieves a lower \ac{RMSE} since it processes the raw received digital signal without physical restrictions stemming from the specific hardware, hence applying unconstrained mathematical operations (e.g., arbitrary matrix multiplications), while requiring a fully digital implementation (1600 \ac{RF} chains \emph{vs.} 2 \ac{RF} chains of the \ac{NL-SIM}). In contrast, the \ac{SIM} and \ac{NL-SIM} computations are strictly limited by the laws of \ac{EM} wave propagation, as per~\eqref{eq:intralayprop}, and the specific metasurface geometry, which limits its generalization capabilities when compared to a software-based \ac{NN}. Furthermore, the \ac{ML} estimator attains the smallest localization error, i.e., approximately $8\,\mathrm{mm}$, albeit at the cost of extremely high hardware complexity ($1600$ \ac{RF} chains) or prohibitive computational search requirements. In contrast, the proposed \ac{NL-SIM} achieves comparable accuracy with a one-shot estimation while substantially reducing the number of antenna elements.

\section{Conclusions}

This work introduced the \ac{NL-SIM} architecture by enriching stacked metasurfaces with passive \ac{NL} cells, enabling multifunctional EM-domain signal processing. A rigorous \ac{EM}-based framework was developed, capturing both inter-layer propagation and the constraints imposed by realistic \ac{RF} \ac{NL} responses, alongside the proposal of a simple diode-based implementation for passive \ac{NL} operations. 

The numerical study on near-field localization confirms the potential of this architecture. \ac{NL-SIM} achieves accuracy comparable to far more complex fully digital front-ends while relying on a limited number of \ac{RF} chains. The inclusion of \ac{NL} provides a clear performance advantage over linear \ac{SIM}, particularly when applied at the final layer, while deeper stacks further enhance accuracy. Interestingly, even static random \ac{NL} cells perform close to fully trainable designs, indicating that effective low-power, passive implementations are feasible. 

Overall, these findings demonstrate that embedding \ac{NL} directly in the \ac{EM} domain significantly increases the expressive power of metasurface-based receivers, enabling over-the-air regression tasks without additional analog or digital processing.

\bibliographystyle{ieeetr}
\bibliography{references}
\end{document}